\documentclass{sig-alternate-2013}

\usepackage{url}
\usepackage{graphicx}
\usepackage{epstopdf}
\usepackage{epsfig}
\usepackage{mdwlist}
\usepackage{flushend}
\usepackage{array}
\usepackage{bm}
\usepackage{algorithm}
\usepackage{algpseudocode}
\usepackage{multirow}
\usepackage{colortbl}
\usepackage{balance}
\usepackage[T1]{fontenc}
\usepackage{tabularx}
\usepackage{tabulary}
\usepackage{subcaption}
\usepackage{array}
\usepackage{tensor}
\usepackage{enumitem}

\setdescription{noitemsep} 
\newcolumntype{P}[1]{>{\centering\arraybackslash}p{#1}}

\DeclareMathOperator*{\argmax}{argmax}

\newcommand{\action}{a}
\newcommand{\actions}{\vec{\mathbf{a}}}
\newcommand{\actionspace}{\mathcal{A}}
\newcommand{\utilityS}{U_S}
\newcommand{\utilityD}{U_D}
\newcommand{\userone}{\text{{\tt user}}_1}
\newcommand{\usertwo}{\text{{\tt user}}_2}
\newcommand{\observation}{o}
\newcommand{\observations}{\vec{\mathbf{o}}}
\newcommand{\observationSpace}{\mathcal{O}}
\newcommand{\relevance}{r}
\newcommand{\relevancies}{\vec{\mathbf{r}}}
\newcommand{\relupdate}{\tau}
\newcommand{\pathfunc}{\omega}

\usepackage{mathtools}

\makeatletter
\def\@copyrightspace{\relax}
\makeatother

\begin{document}
\setlength{\pdfpagewidth}{8.5in}

\setlength{\pdfpageheight}{11in}

\title{Dynamic Information Retrieval: Theoretical Framework and Application}

\numberofauthors{1}
\author{
\alignauthor
Marc Sloan and Jun Wang\\
\affaddr{Department of Computer Science}\\
\affaddr{University College London}\\
 \email{M.Sloan@cs.ucl.ac.uk, J.Wang@cs.ucl.ac.uk}
 \alignauthor }

\maketitle 
\begin{abstract}

Theoretical frameworks like the \emph{Probability Ranking Principle} and its more recent \emph{Interactive Information Retrieval} variant have guided the development of ranking and retrieval algorithms for decades, yet they are not capable of  helping us model problems in \emph{Dynamic Information Retrieval} which exhibit the following three properties; an observable user signal, retrieval over multiple stages and an overall search intent. In this paper a new theoretical framework for retrieval in these scenarios is proposed. We derive a general dynamic utility function for optimizing over these types of tasks, that takes into account the utility of each stage and the probability of observing user feedback. We apply our framework to experiments over TREC data in the dynamic multi page search scenario as a practical demonstration of its effectiveness and to frame the discussion of its use, its limitations and to compare it against the existing frameworks. 

\end{abstract}
\vspace{-8pt}
\noindent
\category{H.3.3}{Information Search and Retrieval} Relevance feedback; Retrieval Models; Search process;
\vspace{-12pt}
\noindent

\section{Introduction}

The theoretical frameworks that underpin research in Information Retrieval (IR) are based on abstract models of user benefit. For instance, the loss function defined in the classic \emph{Probability Ranking Principle} (\emph{PRP})~\cite{JD:1977:Robertson:PRP} leads to justification for the simplest and most powerful ranking rule in IR; ranking documents in decreasing order of their probability of relevance. A recent counterpart is the \emph{Probability Ranking Principle for Interactive Information Retrieval} (\emph{IIR-PRP})~\cite{DBLP:journals/ir/Fuhr08}, which relaxes the independence assumption in the \emph{PRP}'s model to take into account non-linear decision making. For example, document dependence is a key element in IR diversification that is not handled by the \emph{PRP}~\cite{Wang09portfoliotheory}. These models deal with traditional ad hoc query ranking and retrieval. Yet search tasks are complex and often exploratory, being comprised of multiple stages of retrieval with information needs specialized or generalized over time~\cite{white2009exploratory}. Throughout, a user may broadcast signals of search intent that can help an IR system to improve retrieval. Examples include query reformulation in session search~\cite{my-ijr-work} and item ratings in collaborative filtering~\cite{Jambor:2012:UCT:2187836.2187839}. The described models are not capable of representing search tasks that operate over multiple stages nor can they incorporate user feedback. 


These types of problems belong to the area of IR research known as \emph{Dynamic Information Retrieval} (\emph{DIR}), which we define as exhibiting three characteristics: user feedback, temporal dependency and an overall goal. In this paper we present \emph{DIR} as a natural progression in IR research complexity; where early research concerned \emph{static} problems such as ad hoc retrieval, which gave way to \emph{interactive} tasks such as those incorporating relevance feedback~\cite{Rocchio}, finally leading to \emph{dynamic} systems where tasks such as ranking for session search are optimized~\cite{grace-ecir}. 

From this progression we mathematically formulate a generalized framework that models the expected benefit to a user of completing a \emph{DIR} task. This benefit is represented as a recursive utility function that is goal oriented and adaptive over time. The components of this utility function represent the three \emph{DIR} features: the likelihood of user feedback, a probability of relevance model conditioned on this feedback and an individual stage utility function. The optimization of this recursive utility leads to an optimal policy of actions dependent on user interactions in the dynamic setting. 


This utility is shown to be a form of Bellman equation~\cite{Bellman:2003:DP:862270}, the framework an instantiation of a \emph{Partially Observable Markov Decision Process} (\emph{POMDP})~\cite{Sondik:1978} and also a generalization over existing research in \emph{DIR}~\cite{grace-ecir,Jin:2013:IES:2488388.2488446}. The components of the utility can also be linked to the cost-benefit parameters of the \emph{IIR-PRP} and the discount-gain functions found in session-based metrics such as \emph{sDCG}~\cite{Kanoulas:2011:EMS:2009916.2010056}. The structure of the utility function and its links to these areas of research give us interesting insights into the behavior of the function in \emph{DIR} problems. These insights, such as how the quality and diversification of rankings vary over multiple stages, are supported by our experiments performed using a specific application of the \emph{DIR} utility function over TREC data. Our experimental setting is the multi-page search scenario of choosing optimal rankings to display over several search pages for a fixed query~\cite{Kim:2013:UPI:2541176.2505663,Jin:2013:IES:2488388.2488446}, a simplified \emph{DIR} problem. As well as being a demonstration of the implementation of each of the components that make up the utility, practical aspects such as the computational complexity are also explored. 

Thus, through its supporting theory (Section~\ref{DIR}) and application (Section~\ref{application}), we establish our dynamic utility function (Section~\ref{opt-DIR}) as a new theoretical framework for the modeling of dynamic information retrieval problems.



\section{Comparison of IR Frameworks}
\label{DIR}

\begin{figure*}[t!]
        \centering
                \caption{An example illustration of document ranking and relevance feedback using the vector space model for query $Q_1 = $ {\tt apple}. Documents are given as points over two term frequency axes, {\tt computer} and {\tt fruit}, and can belong to one of three subtopics {\tt apple fruit}, {\tt apple logo} and {\tt apple computer}. The distance between $Q$ and each document is inversely proportional to its relevance $r$. The documents ranked for $Q_1$ or its reformulation $Q_2$ are contained in each circular shape $\actions$, whose area could be thought of as the static utility $\utilityS(\actions, \relevancies)$, or $\utilityD$ the combined area of actions across stages 1 and 2.}
        \label{fig:dir-example}
        \begin{subfigure}[t]{0.238\textwidth}
                \includegraphics[width=\textwidth]{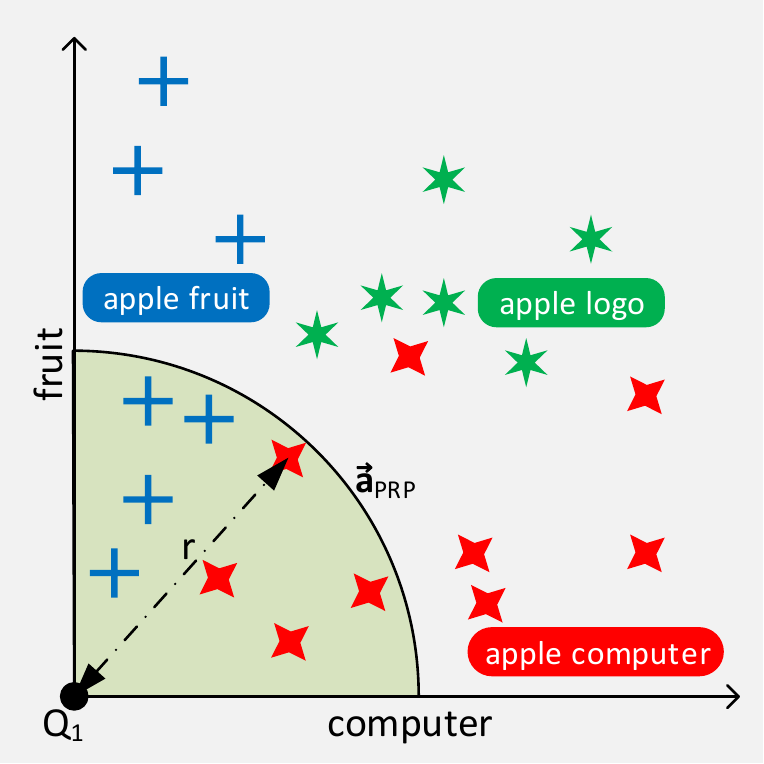}
                \caption{Static IR: Documents within the ranking $\actions_{\text{PRP}}$ are shown to the user for query $Q_1$, but do not cover all subtopics. Optimally ranking using the \emph{PRP} results in choosing those documents with the highest relevance.}
                \label{fig:dir-example-a}
        \end{subfigure}%
        ~ 
        \begin{subfigure}[t]{0.238\textwidth}
                \includegraphics[width=\textwidth]{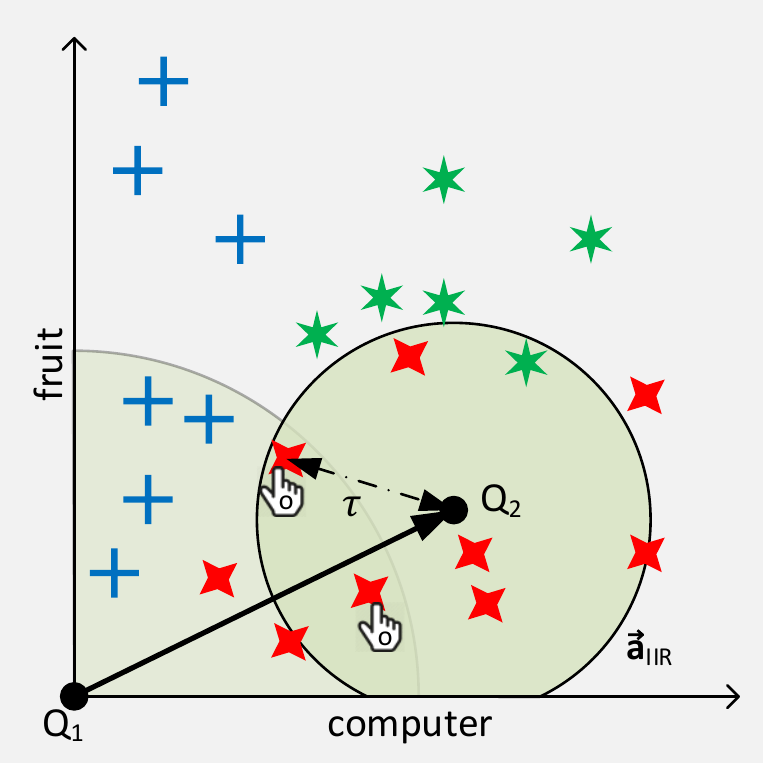}
                \caption{Interactive IR: After relevance feedback is observed (the two click observations $\observation$) on the static ranking in Fig.~\ref{fig:dir-example-a}, $Q_1$ is modified to $Q_2$. Document relevance for $Q_2$ is now defined by $\relupdate$ and the new interactive ranking given by documents in $\actions_{\text{IIR}}$.}
                \label{fig:dir-example-b}
        \end{subfigure}
        ~ 
        \begin{subfigure}[t]{0.238\textwidth}
                \includegraphics[width=\textwidth]{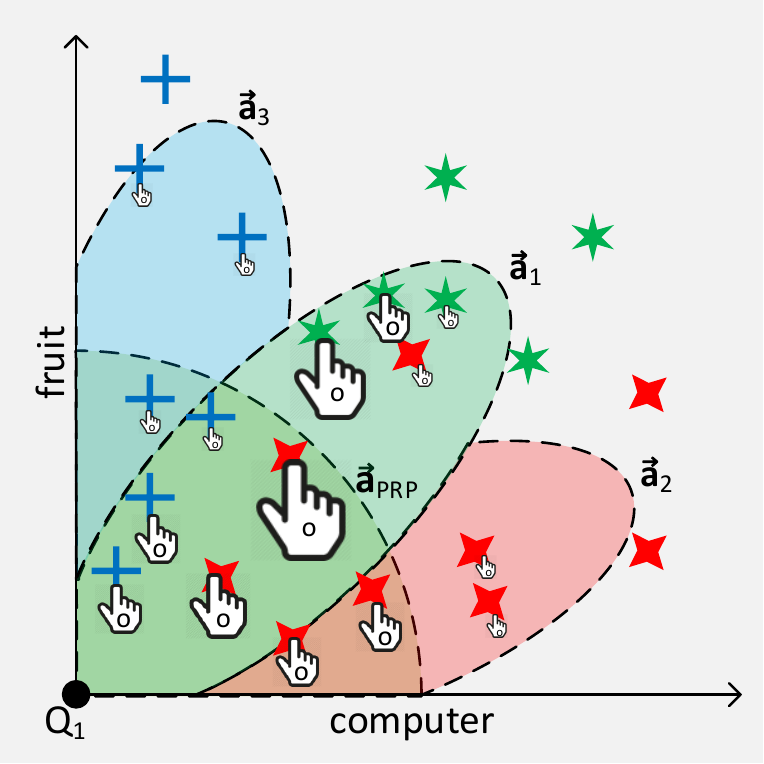}
                \caption{Dynamic IR: Four potential rankings $\actions_1$, $\actions_2$,  $\actions_3$ and $\actions_{\text{PRP}}$ and their observation probabilities (shown as click observations with likelihood relative to size) for $Q_1$ are explored to find the optimal ranking action for both stages.}
                \label{fig:dir-example-c}
        \end{subfigure}
                ~ 
        \begin{subfigure}[t]{0.238\textwidth}
                \includegraphics[width=\textwidth]{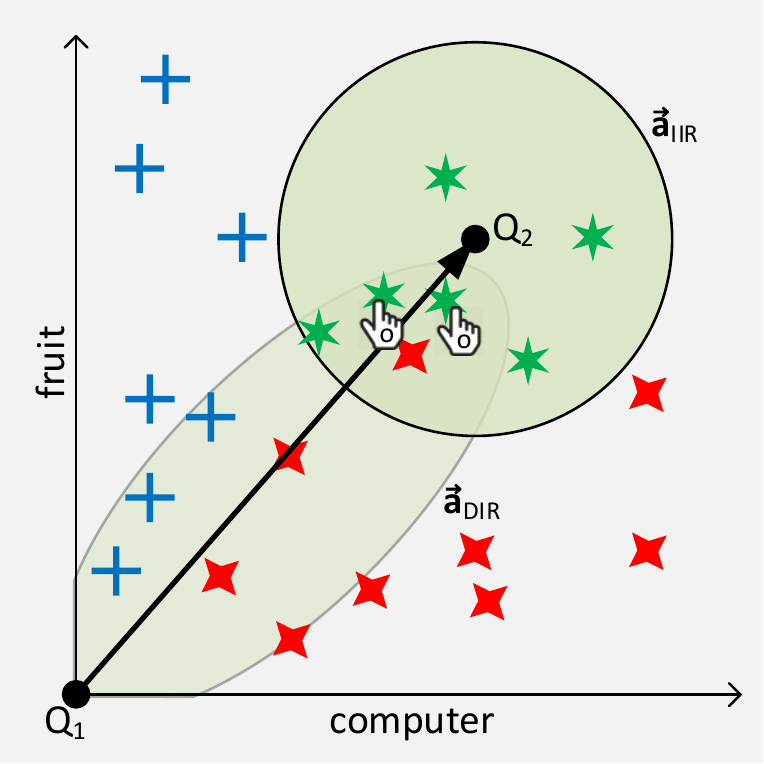}
                \caption{Optimal Solution: Action $\actions_1$ is chosen as the optimal stage 1 ranking $\actions_{\text{DIR}}$ as it diversely contains documents from all subtopics. As a result, the ranking $\actions_{\text{IIR}}$ for $Q_2$ is more accurately modified after observing interactive feedback.}
                \label{fig:dir-example-d}
        \end{subfigure}
\end{figure*}

%

Before setting up our framework for dynamic information retrieval, we consider \emph{DIR} in the context of existing static and interactive theoretical IR frameworks in order to mathematically identify those features that distinguish it. 

\subsection{Static IR Framework}


\emph{\textbf{Definition:}} A static IR framework is one which models single user interactions, or else multiple \emph{independent} interactions of different search intents. A typical application would be an ad hoc ranking and retrieval system.

The objective of a static system is to choose an \textbf{action} $\action$ (or sequence of actions $\actions = \langle \action_1,\action_2,\ldots \rangle$), each of which has an associated \textbf{probability of relevance} $\relevance$ (or $\relevancies$ for a sequence of actions) that maximizes some \textbf{static utility function} $\utilityS(\action, \relevance)$. The action represents a choice that can be made by the system and belongs to some action space $\actionspace$. For example, $\action$ may be a query suggestion to display to a user, or $\actions$ the ranking order of a set of documents for retrieval. The utility function gives value to the action based on its probability of relevance by modeling the benefit of the action to the user. Utilities such as expected DCG and MAP~\cite{WANG:2010:SAO:1835449.1835489} are examples from document retrieval, rewarding the ranking of relevant documents at high ranking positions. 

\subsubsection{Probability Ranking Principle (PRP)}

The \emph{PRP} defines $\utilityS(\action, \relevance)$ as a loss minimizing function across pairs of documents, which is optimized when ranking documents in decreasing order of probability of relevance (under document independence assumptions)~\cite{JD:1977:Robertson:PRP}. Nonetheless, in instances where result diversity is important, it can be shown that \emph{PRP} is no longer optimal \cite{Wang09portfoliotheory}. We illustrate this with our example in Fig.~\ref{fig:dir-example-a}. Here, we represent a simplified vector space model for ranking and retrieval using a graph over two term axes. In this case, the query is {\tt apple}, an ambiguous term that can describe three subtopic search intents. Those documents within the ranking $\actions_{\text{PRP}}$ for the query are retrieved (analogous to ranking under the \emph{PRP}), and as we can see, in this case only two subtopic preferences are captured. Over a population of users, those seeking information on the {\tt apple logo} subtopic would be dissatisfied. 

We can capture this probabilistically by supposing that we have two classes of users, $\userone$ and $\usertwo$, where $\userone$ has twice as many members as $\usertwo$.  Users in the $\userone$ class are satisfied with the {\tt apple logo} and {\tt apple computer} subtopics, but not {\tt apple fruit}, while those in the $\usertwo$ class are only satisfied with the {\tt apple fruit} subtopic. Our action space here is the set of subtopics which we denote $\{\action_1 = \text{{\tt apple logo}}, \action_2 = \text{{\tt apple computer}}, \action_3 = \text{{\tt apple fruit}}\}$ and our goal is to choose the best ranking of subtopics. 

If we set $R_{\action_k} = 1$ if $\action_k$ is relevant, and $r_{\action_k} = P(R_{\action_k} = 1)$, then we have $r_{\action_1} = \frac{2}{3}, r_{\action_2} = \frac{2}{3}$ and $r_{\action_3} = \frac{1}{3}$. According to the \emph{PRP}, we should rank in decreasing order of the probability of relevance, giving us the ranking sequence $\actions_{\text{PRP}} = \langle\action_1, \action_2,  \action_3\rangle$. However, intuitively this is not optimal because users belonging to $\usertwo$ have to reject two subtopics before reaching their preference~\cite{cooper1971inadequacy}. We can explain this mathematically by studying the optimization of the diversity-encouraging metric Expected Search Length. One can also derive the same conclusion analogously using the equivalent Expected Reciprocal Rank or $k$-call at $n$ measures~\cite{Chen:2006:LMP:1148170.1148245}. In this scenario, $\utilityS(\actions, \relevancies) = E[L]_{\actions}$ which is the summation of all possible search lengths $L$ weighted by their respective probabilities, given as
\begin{equation}
  E[L]_{\actions} = \sum_i \big((i-1) P(R_1=0,\ldots,R_{i-1}=0, R_i=1) \big) \nonumber 
\end{equation} 
where $R_i$ is the relevance of the subtopic at rank position $i$. When assuming subtopics are independent, i.e. $P(R_1=0,\ldots,R_i=1) = P(R_{1}=0)\ldots P(R_{i-1}=0) P(R_i=1) $
the expected search length for ranking $\actions_{\text{PRP}} $ is
 \begin{align*}                                                                                                                                                                                                                                  &E[L]_{\actions_{\text{PRP}} } = \ 0\!\cdot\!r_{\action_1}+1\!\cdot\!r_{\action_2}(1\!-\! r_{\action_1})+2\!\cdot\! r_{\action_3}(1\!-\!r_{\action_2})(1\!-\!r_{\action_1})\\
&\ \ \ \  = \ 0\!\cdot\!(2/3) + 1\!\cdot\! (2/3)(1/3) + 2\!\cdot\! (1/3)(1/3)(1/3) = \mathbf{8/27}
\end{align*} 
and for a diversified ranking $\actions_{\text{DIV}}= \langle \action_1, \action_3, \action_2\rangle $ the expected search length is
\begin{align*}
  E[L]_{\actions_{\text{DIV}}} \!=\!&\  0\!\cdot\!(2/3)\!+\!1\!\cdot\! (1/3)(1/3)\! +\! 2\!\cdot\! (2/3)^2(1/3)  \!=\!\mathbf{11/27 }
  \end{align*}

Thus, in this case the \emph{PRP} ranked documents have a shorter expected search path than the diversified ranking. Here, the \emph{PRP} does lead to the optimal ranking under the independence assumption, but when we remove it this is no longer the case. To see this, we recalculate the expected search length for $\actions_{\text{PRP}}$ and $\actions_{\text{DIV}} $ but this time without the independence assumption:  
\begin{align*}                                                                                                                                                                                                                                                                                                                                                                                                                                                                                                                                                        E[L]_{\actions_{\text{PRP}} } =&\ 0 \cdot r_{\action_1}+ 1\cdot  P(R_{\action_2}=1, R_{\action_1}=0)) \\
 &\ \ \ \ \ + 2\cdot P(R_{\action_3}=1, R_{\action_2}=0, R_{\action_1}=0)  \\
=&\  0\cdot (2/3) + 1\cdot 0 + 2\cdot (1/3) = \  \mathbf{2/3}\\
E[L]_{\actions_{\text{DIV}}} =& \ 0\cdot (2/3) + 1\cdot (1/3) + 2\cdot 0 =\ \mathbf{1/3}
\end{align*}

Now we find that the diversified ranking $\actions_{\text{DIV}} $ has the shorter expected search length and is thus the optimal ranking, despite the lower probability of relevance for $\action_3$. 

\subsection{Interactive IR Framework}

\emph{\textbf{Definition:}} An interactive IR framework extends a static framework to cover multiple stages of IR. It is \emph{responsive} to feedback from a previous stage but does not anticipate future feedback. 

A \textbf{stage} represents an interaction with the search system that is distinct from other interactions but belongs to the same search task, for example a sequence of impressions in session search. Generally, an IR system will operate over $1 \leq t \leq T$ stages with $T$ being potentially infinite. 

Further to this, an interactive IR framework incorporates user feedback. Feedback is an \textbf{observation} signal $\observation$ (or a sequence of observations $\observations$) in the space $\mathcal{O}$, that is measurable by the search system. These signals may be \emph{explicit} declarations of the relevance of search items (such as a movie rating), or \emph{implicit} interpretations of user actions (such as document clicks). 

The final element of this framework is the \textbf{relevance update function} $\relupdate$ where $\relevance_{t + 1} = \relupdate(\action_t, \relevance_t, \observation_t)$. Thus, the objective function for interactive IR at stage $t + 1$ can now be represented as $\argmax_{\action_{t + 1} \in \actionspace} \utilityS \bigl(\action_{t + 1}, \relupdate(\action_{t }, \relevance_{t }, \observation_{t })\bigr)$.
The relevance update function $\relupdate$ introduces temporal dependency into the framework, without it the objective simply devolves to optimization over the static utility $\utilityS(\action, \relevance)$. This is also the case when finding the optimal first stage action $\action_1$ i.e. when there are not yet any observations. In interactive IR, the optimal action is chosen at each stage as a reaction to the feedback observed in the previous stage and there is no consideration for future utility. 

With these features in mind, we extend the vector space example to the interactive scenario in Fig.~\ref{fig:dir-example-b} by introducing the Rocchio relevance feedback algorithm~\cite{Rocchio} for interactively re-ranking documents. Here, clicked documents in the \emph{PRP} ranked first stage are used as implicit signals of relevance to modify the user's original query $Q_1$ to $Q_2$. Document re-retrieval occurs using $Q_2$, returning documents using updated relevance scores given by $\relupdate$, which is a function of $Q_2$ and thus the original ranking $\actions_{\text{PRP}}$, document relevancies $\relevance$ and observations $\observation$. Nonetheless, even in this interactive framework, a user interested in the {\tt apple logo} subtopic would be dissatisfied with both the $\actions_{\text{PRP}}$ and $\actions_{\text{IIR}}$ rankings due to a lack of documents for the relevant subtopic. 

\subsubsection{Interactive Information Retrieval (IIR)}

For clarification, the area of research traditionally known as \emph{Interactive Information Retrieval} (\emph{IIR}) has an alternative definition to the interactive IR \emph{framework} discussed in this paper, despite the similarity in name. \emph{IIR} research explores the complex sequence of interactions a user may have with a search ranking within the static framework~\cite{DBLP:journals/arist/Ruthven08}, largely motivated by the contradictory results found from conventional Cranfield style evaluation~\cite{Cleverdon68} and observational user studies~\cite{Hersh:2000:BUE:345508.345539}. With the exception of the \emph{IIR-PRP} framework which we cover in more detail in Section~\ref{sec:iir-prp}, for the remainder of this paper any reference to interactive IR instead reflects the framework defined in this paper.



\section{Dynamic IR Theory}
\label{opt-DIR}

A dynamic system is one which is goal-directed and adaptive to its environment. From this definition we can specify three elements that determine whether an IR system is a dynamic one:
\vspace{10pt}
\begin{description}[topsep=0.5ex]
\item [Feedback] An observation signal from the user.
\item [Temporal Dependency] Operation across multiple stages where each stage depends on the previous stage. 
\item [Overall Goal] An objective across all stages. 
\end{description}

\subsection{Dynamic IR Framework}

\emph{\textbf{Definition:}} A dynamic IR framework extends an interactive framework by being responsive to user feedback \emph{and} optimizing for it in advance.

We previously defined systems in the interactive IR framework as exhibiting both feedback and temporal dependency features, but they are only capable of locally optimizing for a single stage at a time. In contrast, the optimization of a dynamic system will find the optimal sequence of actions for all future interactions. A result of this is that the utility of an individual stage may be reduced so that gains can be made in the utility at a future stage. 

Unlike the interactive IR framework, the observation $\observation$ is unknown when evaluating the utility of future stages in the dynamic IR framework. Instead, the \emph{expected} utility can be found by marginalizing the utility function over the space of observations $\observationSpace$. When doing this the \textbf{observation likelihood function} $P(\observation | \action, \relevance)$ must be specified. This gives the {expected} utility
\begin{align}
E[\utilityS(\action, \relevance)] = \sum_{\observation \in \observationSpace} P(\observation | \action, \relevance) U_S(\action, \relevance) \label{eq:expectedutility}
\end{align}

The observation likelihood function is represented visually in Fig.~\ref{fig:dir-example-c}. In dynamic IR, the expected utility of potential first stage rankings (given here as $\actions_1$, $\actions_2$, $\actions_3$ and the optimal static ranking $\actions_{\text{PRP}}$) are calculated by estimating which documents are likely to receive clicks and the effect this has on the utility of future stages. The \emph{PRP} ranking is simply one among many rank actions that can be considered. 


The final component of the \emph{DIR} framework is the \textbf{path discount} function $\pathfunc(t)$. When optimizing over a potentially infinite number of future stages, this helps ensure that a solution exists and also gives greater weight to earlier stage utilities.

By bringing together all of the components described so far, we can define the \textbf{utility function for dynamic information retrieval} as 
\begin{align}
\utilityD(\relevance_t, t) =& \max_{\action_t \in \actionspace}\bigl[\utilityS(\action_t, \relevance_t) + \nonumber \\
&\ \ \ \pathfunc(t)\sum_{\observation \in \observationSpace}P(\observation | \action_t, \relevance_t)\utilityD(\relupdate(\action_t, \relevance_t, \observation), t+1)\bigr] \label{eq:dynamic-utility}
\end{align}
 where $\utilityD(\relevance_T, T) = \max_{\action_T \in \actionspace} \bigl[\utilityS(\action_T, \relevance_T)\bigr]$ is the static optimization of the final stage. Thus, our objective is to find, through backwards induction, the optimal sequence of actions $\actions^* = \langle \action_1, \ldots, \action_T \rangle$ that maximizes the \textbf{dynamic utility} $\utilityD$ given in  Eq.~(\ref{eq:dynamic-utility}). To derive this utility we have simply recursively applied the dynamic utility to the expected utility from Eq.~(\ref{eq:expectedutility}). 
 
Through the maximization of the dynamic utility, in our example in Fig.~\ref{fig:dir-example-d} we find the optimal action for stage 1, which is to diversify the initial ranking so that it retrieves documents belonging to all three subtopics. While this may harm the immediate retrieval utility score, overall the system improves because it can more accurately re-rank results over the subtopic preferences for all users in the next stage. 
 
 

We observe that the eight elements of the \emph{DIR} framework: $\action$, $\relevance$, $\utilityS$, $t$, $\observation$,  $\relupdate$, $P(\observation | \action, \relevance)$ and $\pathfunc(t)$, are also the elements that define a \emph{POMDP}~\cite{Sondik:1978}, and that the dynamic utility function is its corresponding Bellman equation~\cite{Bellman:2003:DP:862270}. Intuitively this makes sense, like a \emph{POMDP} the dynamic IR framework finds an optimal Markovian sequence of actions to maximize a reward (here the static utility) subject to discounting (with $\pathfunc$). The state of the system (the underlying document relevance) is unknown but a belief state (the probability of relevance) is updated according to observations. The key difference from a \emph{POMDP} is that for dynamic IR we do not define a transition probability between states as we assume that the hidden relevance of each document does not change throughout the search task.


\subsection{Framework Analysis}

So far we have described the general framework for dynamic IR but have not addressed the setting of its parameters. Here we analyze each component within the context of dynamic information retrieval. 
\vspace{-5pt}

\paragraph*{Relevance} As with any framework in information retrieval, the overall aim is to retrieve relevant information items and present them to the user. The intrinsic `relevance' of an information item is an unknown quality and the subject of most of the research in IR. In the \emph{DIR} framework any document relevance scoring method can be used. For instance, in our application in the next section we make use of five well established relevance scoring techniques. 

The relevance update function $\relupdate$ is more difficult to define as it depends specifically on the action and observation space of the \emph{DIR} task, for instance in Fig.~\ref{fig:dir-example-b} it depends on the distance of the documents from $Q_2$, which itself depends on $Q_1$, $\actions_{\text{PRP}}$ and its clicked documents. This dependence allows $\relupdate$ to adapt to the hidden relevance preferences of the user over the course of the search process. 

It may not always be clear how to update the relevance score based on a given observation, the most straightforward setting for $\relupdate$ can simply be to set $\relevance_{\action} = 0$ for actions already chosen by the IR system. Because $\relupdate$ enforces the temporal dependency, it is the most important aspect in the dynamic utility because without it the utility is static.  
\vspace{-5pt}
\paragraph*{Actions} The action space is what distinguishes search tasks from one another and it is the size of this space that dictates the complexity of optimizing over the dynamic utility function. For example, the action space in query suggestion or document ranking is potentially infinite whereas the space of available advertisements in an ad selection problem may be small and finite. In our example the action space is any potential grouping of the documents in the 2D term space (four such groupings are shown in Fig.~\ref{fig:dir-example-c}). Along with $\relupdate$, the setting of the static utility $\utilityS$ is important for determining the desirable features of the optimal action sequence $\actions^*$, such as results diversification. 
\vspace{-5pt}
\paragraph*{Observations} The observation space is dependent on the action space, its elements representing the user's response to system actions. Each observation must contain some signal of relevance or search intent, otherwise we would have $\relupdate(\action, \relevance, \observation) = \relupdate(\action, \relevance)$ and lose the temporal dependency. In some cases the value of the observation likelihood is simply $P(\observation | \action, \relevance) = \relevance$, for instance in search tasks where accurate explicit relevance feedback is guaranteed. Otherwise, in most situations the observations will be click-related and thus the observation probability is the probability of click, as is the case in our example (Fig.~\ref{fig:dir-example}).
\vspace{-5pt}
\paragraph*{Stages} Typically, the stages in a \emph{DIR} task will represent distinct interactions occurring in a linear time order. In these cases $\pathfunc(t)$ may take a value between 0 and 1 or be set to a monotonically decreasing function that favorably weights the utility scores of immediate stages. Setting $\pathfunc(1) = 1$ and $\pathfunc(t) = 0$ for $t > 1$ gives us the static and interactive scenarios. 

Alternatively, a non-linear sequence of interactions (or \emph{search path}) can be modeled as Yang and Lad did with their session-based utility function~\cite{yangLad}. For instance in session search, a search path represents a particular sequence of documents examined by the user and the query reformulations made. For our framework, the stage $t$ may instead represent a specific search path, and so $\pathfunc(t)$ could be interpreted as the likelihood of this path rather than an explicit discount, penalizing improbable search paths and rewarding likely ones. 

The time horizon $T$ dictates the number of advance stages to optimize for. A large time horizon will lead to explorative action strategies that benefit later stages. In our experiments, we set $T=2$ so that we only consider exploitative optimizations for the immediate next stage. 
\vspace{-5pt}
 \paragraph*{Dynamic Utility} Through the recursive evaluation of the utility function we not only learn the optimal sequence of actions to make in the dynamic system, but we also learn the optimal action for each possible observation at each stage. If we were to store these in a lookup table ahead of deployment, then the dynamic system would be immediately responsive to user feedback and able to cater to a population of users. Nonetheless, the construction and storage complexity of such a table may prove intractable. We also note that the static utility $\utilityS$ may be set as the dynamic utility function $\utilityD$ of a nested subproblem in the search task. For example, the utility of choosing an optimal r\emph{anking of documents} may be embedded in the utility for determining an optimal \emph{sequence of rankings} for a user in a session, which itself may be defined within the context of modeling a user's \emph{topic preference} from search sessions in their search history. 
 
\subsection{Links to Existing Work}

Building on our analysis of the \emph{DIR} framework, here we identify links between its components and other related work in \emph{IIR} and session search. 

\subsubsection{IIR-PRP}
\label{sec:iir-prp}

The \emph{IIR-PRP}~\cite{DBLP:journals/ir/Fuhr08} is a framework designed for interactive IR in the traditional sense. The objective function in \emph{IIR-PRP} balances the costs and benefits of choosing actions within a sequence and bears some similarities to our dynamic utility function. Nonetheless, by lacking any form of user feedback or temporal dependency, we do not describe the model as interactive or dynamic, and as already discussed, within the terminology used in this paper this means that it is actually a static method. 

By mapping our notation onto the \emph{IIR-PRP} objective function, we get
\begin{align}
U_{\text{IIR}}(\actions, \relevancies) =& \sum_{i = 1}^M  \biggl[\prod_{j = 1}^{i - 1} (1 - \relevance_j)\biggr] \times \nonumber \\
&\ \ \ \ \ \ \  \biggl(\omega(i)+ \relevance_i \sum_{\observation \in \observations}P( \observation| \action_i, \relevance_i)\utilityS(\action_i, \relevance_i)\biggr)\label{eqn:iir-prp}
\end{align}
evaluated over a sequence of $M$ actions (usually a ranking of documents). Eq.~(\ref{eqn:iir-prp}) is a generalization of the formula originally defined, where we recognize that the cost and benefit parameters are simply a utility value, that the probability of whether a user continues searching or not is the observation likelihood, and that the cost of reaching a specific action is the path discount. 

Further to this, the \emph{IIR-PRP} defines a simple ranking rule according to this utility model, ranking documents in order of decreasing value of function $\varrho(\action, \relevance)$, which balances the costs and benefits of choosing an action as well as its probability of relevance. We can define $\varrho$ in our setting as
\begin{align}
\varrho(\actions_{1\ldots i}, \relevancies_{1\ldots i}, \pathfunc) = U_S(\actions_{1\ldots i}, \relevancies_{1\ldots i}) - \frac{\omega}{r_{\action_i}}\prod_{j = 1}^{i - 1}(1 - r_{\action_j}) \label{eq:iir-prp-mps}
\end{align}
where the utility of adding an action $\action_i$ to an existing sequence is countered by the path discount and the probability of not finding previous actions relevant. We implement this in our algorithm for \emph{IIR-PRP} in Algorithm~\ref{algo-iir-prp-mps} in the next section. 

 \subsubsection{Session-Based Utility}
 
 There have been recent advances in the modeling of user benefit across queries in search sessions. This is in recognition of the fact that ad-hoc retrieval often occurs over multiple queries in a session~\cite{white2009exploratory}, with one study finding that 32\% of search sessions consisted of at least three queries~\cite{Jansen:2005:TCA:1059467.1059470}. A simple approach has been to extend discount-gain metrics such as DCG and Average Precision, typically associated with static retrieval, across multiple stages. Using our terminology, a discount-gain function for a single stage has the form $\sum_{i = 1}^M  \pathfunc(i)\utilityS(\action_i, \relevance_i) $. For example, in the DCG metric the setting would be $\pathfunc(i) = \frac{1}{\log_2 (i + 1)}$ and $\utilityS(\action_i, \relevance_i) = 2^{\relevance_{\action_i}} - 1$. For the session-DCG (sDCG)~\cite{Jarvelin:2008:DCG:1793274.1793280} metric, a single layer of recursion is introduced, where 
 \begin{align}
 sDCG = \sum_{t = 1}^T  \pathfunc(t)U(\actions_t, \relevancies_t)  \label{eq:sDCG}
\end{align}
and the discount and gain functions are set as $\pathfunc(t) = \frac{1}{\log_{2t}(t + 1)} $ and 
\begin{align*}
U(\actions_t, \relevancies_t) = DCG(\actions_t, \relevancies_t) = \sum_{i = 1}^M\underbrace{\frac{1}{\log_2(i + 1)}}_{\pathfunc(i)} \times (\underbrace{2^{\relevance_{\action_{ti}}} - 1}_{\utilityS(\action_{ti}, \relevance_{ti})})
 \end{align*}
 
 Here, the stages operate across a linear sequence of search rankings. In the session-Average Precision (sAP) metric, the path of interaction taken by the user is unknown and so the metric function marginalizes over the space of all such paths to find the expected sAP~\cite{Kanoulas:2011:EMS:2009916.2010056}.

\section{Application}
\label{application}

 So far we have formulated a theoretical framework for dynamic IR and derived the dynamic utility function $\utilityD$ given in Eq.~(\ref{eq:dynamic-utility}). In this section we apply this framework to the multi page search problem in  \emph{DIR}. In doing so we demonstrate functional settings for the elements in the framework and their implementation in a workable algorithm, which gives us useful insight into the practical limitations of optimizing over $\utilityD$. We compare our algorithm against \emph{PRP} and \emph{IIR-PRP} based approaches in experiments using TREC data and also investigate static and interactive variants of our objective function. Through this we gain understanding of the effect that dynamic utility optimization has on the quality and diversity of rankings in multi page search.

\subsection{Multi Page Search Problem}

The \emph{Multi Page Search} (MPS) scenario concerns the ranking of documents over multiple pages of search results~\cite{Kim:2013:UPI:2541176.2505663,Jin:2013:IES:2488388.2488446}. MPS typically models exploratory search queries which are more likely to lead to multi-query sessions and multi-page searches~\cite{white2009exploratory} (with one study finding that 27\% of all searches occur over multiple pages~\cite{Jansen:2006:WSW:1138797.1138813}). In this scenario documents are retrieved for a single query, ranked and then segregated into pages of $M$ documents. On each page, a user may examine and click on documents. We assume that the user will return to the results page and move onto the next page and we define a threshold of $T$ pages which the user will search over. The goal in MPS is to create rankings of relevant documents across $T$ pages. For the pages following the first, document clicks can be used to personalize search rankings, a situation analogous to our example in Fig.~\ref{fig:dir-example}. 

We chose this particular problem to apply our dynamic IR framework to as it exhibits the following beneficial features: 1) it is a \emph{DIR} problem that is familiar and easy to define, 2) it is a simple IR scenario where we only have to consider a single query and a single set of documents, 3) we can use existing ad hoc ranking and retrieval research to find suitable implementations for the \emph{DIR} framework components, 4) we can readily use TREC data collections and relevance judgments to evaluate our algorithms, and 5) it is naturally translatable to the \emph{PRP} and \emph{IIR-PRP} frameworks. A similar analysis of \emph{DIR} in the session search scenario was also conducted by Luo et al.~\cite{grace-ecir}

In this scenario, each page of search represents a stage in our framework, with $T$ the threshold number of pages. We nominally set $T=2$ although a larger number of pages is feasible and has been studied by Jin et al.~\cite{Jin:2013:IES:2488388.2488446}. The action sequence $\actions_t = \langle \action_{t1}, \ldots, \action_{tM}\rangle$ represents the ranking of documents for ranks 1 to $M$ on page $t$. Before we can fully implement Eq.~(\ref{eq:dynamic-utility}), we must first define each of its functional components in the context of MPS.

\subsubsection{Expected DCG}

The static utility in multi page search is a measure of the quality of the ranking of documents on each page. As with ad hoc ranking and retrieval, we can evaluate this using a metric such as DCG, MAP or ERR. In the absence of relevance judgments, we can instead find the \emph{expected} metric value which uses probabilities of relevance instead~\cite{WANG:2010:SAO:1835449.1835489}. In our application we set the static utility as the expected DCG function, given by
\begin{align}
\utilityS(\actions_t, \relevancies_t) = \sum_{i = i}^M \frac{2^{\relevance_{\action_{ti}}} - 1 + 2^{\relevance_{\action_{ti}} - 1}\log^2(2)Var[\relevance_{\action_{ti}}]}{\log(i + 1)} \label{eq:expected-dcg}
\end{align}
\\\\This utility also takes into consideration the variance of the document's probability of relevance. 

\subsubsection{Examination Hypothesis}

In multi page search our observations are document clicks, which we regard as an implicit signal of the relevance of a document to the user. Thus, we can utilize the clicks from previous search pages to update our probability of relevance model and personalize the document rankings for future pages. 
 
 For a ranking of $M$ documents, the observation space $\observationSpace$ in MPS for a particular page is the combination of binary click events for each document in the ranking. We denote this as the observation vector $\observations = \langle \observation_1,\ldots, \observation_M\rangle$ where $\observation \in \{0, 1\}$. We could na\"{\i}vely  set the observation likelihood to the uniform distribution  $P(\observation | \action, \relevance) = \frac{1}{|\observationSpace|}$ but eye-tracking studies tell us that this is not the case. Instead, the probability of a click occurring on a ranked document is dependent on not only its probability of relevance but also its rank position, amongst other variables~\cite{Joachims:2007:EAI:1229179.1229181}. The probabilistic modeling of user clicks is an extensive area of IR research and in our application we use the simplest model, the \emph{Examination Hypothesis} model~\cite{clickmodels}.

This model supports the eye tracking research by inferring that the probability of a click on a document in a ranked list is equal to the product of its probability of relevance and the bias of its rank position. Thus, the probability of a sequence of clicks is given by 
 \begin{align}
P(\observations | \actions_t, \relevancies_t) = \prod_{i = 1}^{M}(b_i \relevance_{\action_{ti}})^{\observation_i}(1 - b_i \relevance_{\action_{ti}})^{1 - \observation_i} \label{eq:clickModel}
\end{align}
where $b_i$ is a rank bias parameter. In our implementation we set Eq.~(\ref{eq:clickModel}) as our observation likelihood function and set $b_i = \frac{1}{\log (i + 1)}$ which is the discount value used in our expected DCG utility.

 
\subsubsection{Conditional Multivariate Gaussian Distribution}

Once we have a sequence of click observations for a ranking of documents, we can update the probability of relevance distribution for the remaining documents. Here, we achieve this by defining the distribution of all the probabilities of relevance for all documents in the collection as a multivariate Gaussian distribution $R \sim \mathcal{N}(\relevancies, \Sigma) $, where $R$ is their collective random variable, $\relevancies$ the vector of mean relevance scores and $\Sigma$ the covariance matrix over the documents. $\relevancies$ may be set as any relevance score and $\Sigma$ may be set using document similarity or other correlation scores~\cite{Jin:2013:IES:2488388.2488446}. If $\relevancies$ represents a probability of relevance, then we can set the distribution as a \emph{truncated} multivariate Gaussian bounded between 0 and 1. If it is not possible to define the distribution of a relevance score, then the distribution of the mean of multiple relevance scoring techniques can be derived, resulting in an approximately Gaussian distribution that incorporates multiple signals of relevance. It is this approach we take in our experiment, where we set $\relevancies$ and $\Sigma$ as the means and variances of the retrieval scores from five well known techniques, with the diagonal elements from $\Sigma$ used as variance values for our utility calculation in Eq.~(\ref{eq:expected-dcg}). 

Modeling the relevance distribution in this way allows us to conditionally update the probabilities of relevance $\relevancies$ based on our click observations. For a given rank action $\actions_t$ (which includes both clicked and non-clicked documents in the ranking), we denote the remaining non-ranked documents as $\backslash\actions_t$ and partition our distribution parameters as
\begin{align*}
\relevancies = \begin{bmatrix}
       \relevancies_{\backslash \actions_t}    \\
       \relevancies_{\actions}
     \end{bmatrix}\ \ \ \ 
\Sigma = \begin{bmatrix}
       \Sigma_{\backslash \actions_t \backslash \actions_t} & \Sigma_{\backslash \actions_t\actions_t}          \\
       \Sigma_{ \actions_t\backslash \actions_t} & \Sigma_{ \actions \actions}
     \end{bmatrix}
\end{align*}
We can then update the mean relevance scores and covariance matrix for non ranked documents using the formulae
\begin{align}
\relevancies_{\backslash\actions_t} &= \relevancies_{\backslash\actions_t} + \Sigma_{\backslash \actions_t\actions_t}  \Sigma_{ \actions \actions}^{-1}(\observations -\relevancies_{\actions_t} ) \label{eq:tau-func} \\
\Sigma_{\backslash \actions_t\backslash \actions_t} &= \Sigma_{\backslash \actions_t\backslash \actions_t} - \Sigma_{\backslash \actions_t\actions_t}  \Sigma_{ \actions_t \actions_t}^{-1}\Sigma_{ \actions_t\backslash \actions_t} \nonumber
\end{align}
and observations $\observations$. Thus, for given actions and observations, we can use the functions above to define a new conditional multivariate Gaussian distribution of the probability of relevance of the remaining documents, given as $
R_{t + 1} \sim \mathcal{N}(\relevancies_{\backslash\actions_t}, \Sigma_{\backslash \actions_t\backslash \actions_t} | \actions_t, \observations)$. For the multi page search setting, we define $\relupdate(\action, \relevance, \observation) $ as the relevance update function in Eq.~(\ref{eq:tau-func}). 

\subsubsection{Geometric Discount}

The final component required for our application is the discount function $\pathfunc(t)$. In the multi page scenario, we measure the utility of a linear sequence of document rankings rather than the path-based behavior of users. As such, we adopt the simple discount used in a POMDP, setting $\pathfunc = \lambda$ (which is effectively setting it as the geometric discount $\omega(t) = \lambda^{t-1}$ due to the recursion of the dynamic utility). 

Here, we can consider $\pathfunc(t)$ as the probability of the user visiting page $t$. When $\lambda = 0$, we assume only the first page will be visited, and when $\lambda = 1$ all pages are equally likely and given equal weight. The optimal setting for $\lambda$ will vary depending on the type of searches being performed as well as the corpus and quality of results.

\subsection{DIR-MPS}

Now that we have defined each of the functional components of the dynamic IR framework for the multi page search scenario, we present the DIR-MPS algorithm in Algorithm~\ref{algo-dir-mps}. This algorithm is a direct implementation of the recursive utility function $\utilityD$ in Eq.~(\ref{eq:dynamic-utility}) that determines the optimal sequence of document rankings to display for each page. 

It is worth noting that Algorithm~\ref{algo-dir-mps} and the described settings for the \emph{DIR }framework elements are one such instantiation of the framework in the multi page search scenario. Our motive in this section is not to develop a state of the art new ranking technique but rather to demonstrate the application of the framework to a \emph{DIR} problem.

\begin{algorithm}[!t]
\caption{The DIR-MPS Algorithm}
\label{algo-dir-mps}
\begin{algorithmic}
\Function{DIR-MPS}{$t, \relevancies, \actionspace$}
	\If{$t = T + 1$} \Return $[0, \langle \rangle]$ \EndIf
	\State $\actions_{t}^* = \langle \rangle; \actions^*_{t+1} = \langle \rangle$
	\Loop{$\; i \leftarrow 1$ to $M$} \Comment{Sequential Ranking Decision}
		\State $\actions = \actions_{t}^*; u^* = 0$
		\ForAll{$\ a \in \actionspace\backslash \actions\ $} 
			\State $\actions_t = \langle \actions, a\rangle$ 
			\State $u_t = \utilityS(\actions_t, \relevancies_t)$ \Comment{Eq.~(\ref{eq:expected-dcg})}
			\ForAll{$\observations \in \observationSpace$} 
				\State $\relevancies_{t + 1}= \tau(\actions_t, \relevancies, \observation)$ \Comment{Eq.~(\ref{eq:tau-func})}
				\State $[u_{t+1}, \actions_{t+1}] = \text{DIR-MPS}( t+1,  \relevancies_{t+1}, \actionspace \backslash \actions_t )$
				 \State $u_t = u_t + \lambda \cdot P(\observations| \actions_t, \relevancies_t) \cdot u_{t+1}$\Comment{Eq.~(\ref{eq:clickModel})}
			\EndFor
			\If{$u_t > u^*$} 
				\State $u^*=u_t; \actions_{t}^*=\actions_{t}; \actions_{t+1}^* = \actions_{t+1}$
			\EndIf 
		\EndFor
	\EndLoop
	\State	\Return $[u^*, \langle \actions_{t}^*,\actions_{t+1}^*\rangle]$
\EndFunction
\end{algorithmic}
\end{algorithm}

\subsubsection{Dynamic Utility Approximation}

The DIR-MPS algorithm features a number of approximation techniques that increase its computational efficiency as a way to counteract the inherent complexity of the \emph{DIR} framework (discussed further in Section~\ref{prac-limitations}). Firstly, we reduce the action space of potential rankings by employing a \emph{Sequential Ranking Decision} policy. That is, for each page we find the optimal document to rank at each position one by one. For example, we set $M=1$ and find the document $\action^*$ that maximizes $\utilityS(\action,\relevance_{\action})$. Then we fix this document, set $M= 2$ and find the next in the sequence that maximizes $\utilityS(\langle \action^*, \action \rangle, \relevancies_{\langle \action^*, \action \rangle})$. Continuing in this fashion allows us to find an approximately optimal ranking for a single page, one document at a time, greatly reducing the computational complexity.

A property of the probability distribution given in Eq.~(\ref{eq:clickModel}) also allows us to greatly reduce the observation space. We find that this distribution follows Zipf's law, with a few of the click combinations contributing towards most of the probability mass. In fact, from our experiments we typically found that around 15\% of the combinations contributed to 95\% of the aggregated probability. As such, in our implementation of DIR-MPS we restrict the observation space to only the most probable click combinations that cumulatively sum to 0.95, trading off the potential 5\% inaccuracy for speed. 

Finally, it can be shown that when ranking over a single stage, the expected DCG utility function is maximized when documents are ranked according to the \emph{PRP}~\cite{WANG:2010:SAO:1835449.1835489}. We exploit this to increase the efficiency of our algorithm by ranking the threshold page (where we no longer consider a future temporal dependency) according to the \emph{PRP} over the conditionally updated probabilities of relevance. 
\vspace{15pt}
\subsubsection{IIR-PRP-MPS}

In our experiments we directly compare DIR-MPS against rankings created from the applied \emph{PRP} and \emph{IIR-PRP} ranking rules. With the \emph{Probability Ranking Principle} we can simply rank documents in decreasing order of the probability of relevance across $T$ pages. However, the \emph{IIR-PRP} has no existing direct application to our scenario. Instead, we use our definition of the ranking function $\varrho$ given in Eq.~(\ref{eq:iir-prp-mps}) to create the IIR-PRP-MPS algorithm shown in Algorithm~\ref{algo-iir-prp-mps}.

Here, the sequential ranking rule is also employed to build up an optimal ranking over all pages, one document at a time, by selecting the document that has the highest $\varrho$ value for each rank. Thus, there is some dependency on previously ranked documents, which is not possible in the \emph{PRP}, but like the \emph{PRP} there is no way to take into account user feedback or update the probabilities of relevance. 

\begin{algorithm}[!t]
\caption{The IIR-PRP-MPS algorithm}
\label{algo-iir-prp-mps}
\begin{algorithmic}
\Function{IIR-PRP-MPS}{$M, T, \lambda, \actionspace, \relevancies$}
	\State $\actions^* = \langle \rangle$
	\Loop{$\; i \leftarrow 1$ to $M \times T$}
		\State $\action^* = \argmax_{\action \in \actionspace \backslash \actions^*}\varrho(\langle \actions^*, \action \rangle, \relevancies, \lambda)$ \Comment{Eq.~(\ref{eq:iir-prp-mps})}
		\State $\actions^* = \langle \actions^*, \action^* \rangle$ 
	\EndLoop
	\State	\Return $\actions^* $
\EndFunction
\end{algorithmic}
\end{algorithm}

\subsection{Practical Limitations}
\label{prac-limitations}


The general computational complexity of the optimization of $U_D$ can be shown to be PSPACE-Complete (through its connection to \emph{POMDP}s). For small $T$ and observation and action spaces this can be reasonable, but typically these spaces may be impractically large. 

For example, an IR task such as information filtering or music recommendation may operate over potentially infinite time steps. In these cases the discount factor and threshold $T$ are important. Further, the observation space may not be as well defined as that in our multi page search scenario where $|\observationSpace| = 2^M$, for example, the space of possible reformulations for a query or 2D gaze positions in eye-tracking. Finally, the action space can be difficult to optimize over as is the case with DIR-MPS, where the sequential ranking decision reduces the size of the action space from $O({N!}/{(N - TM)!)}$ to $O(TNM - TM^2)$ for a collection of $N$ documents. Our application serves to demonstrate that such approximations may be needed when working with the \emph{DIR} framework, especially given that the optimization of $U_D$ is not guaranteed to be tractable, and an optimal solution may not exist depending on the particular problem settings.

		
\subsection{Experiment}

To gain insight into our application of the dynamic IR framework in the multi page search scenario, and to compare with the other theoretical frameworks, we conducted an experiment using the WT10g, AQUAINT and ClueWeb09 datasets, the details of which are included in Table~\ref{TBL:Corpus}. We chose these collections as they were designed for evaluating ranking and retrieval algorithms and were easily extended to the multi page problem. The WT10g dataset allowed us to test the theoretical frameworks in the standard ad hoc ranking and retrieval environment. The Robust data consists of difficult to rank ad hoc queries which we hypothesized would be more likely to require several pages of search results. The diversity track data allowed us to test our hypothesis that dynamic optimization leads to increased diversification in ad hoc retrieval. A drawback to using these datasets is that they lack interaction data, which is not needed when optimizing for probable clicks in the \emph{DIR} framework, but important in the interactive setting. 

\begin{table}[t]
  \centering
  \caption{Overview of the three TREC test collections}
  \setlength\extrarowheight{0.1pt}
  \setlength\tabcolsep{3pt} 
  \begin{tabular}{| p{1.55cm} | p{1.9cm} | p{1.58cm} | p{2.25cm} | }
  \hline
Name & Task & \# Docs & Topics \\
\hline  
\hline  
WT10g & TREC 9 Web Track & 1,692,096 & 451-500 \\
\hline
AQUAINT  & Robust 2005 & 1,033,461 & 50 difficult Robust 2004 topics \\
\hline
ClueWeb09 & Diversity Task 2009/10 & 503,903,810 & 1-100 (461 subtopics)\\
\hline
  \end{tabular}
\label{TBL:Corpus}
\end{table}

On each collection we retrieved the top 100 documents for each topic scored using each of the TF-IDF, BM25, Jelinek-Mercer, Dirichlet and Two-Stage language model retrieval methods from the Indri\footnote{\url{http://www.lemurproject.org/indri.php}} search engine. We pooled the documents and subsequently scored them over all the techniques. This gave us an average of 193 ranked documents per topic each with 5 relevance score values. After min-max normalization we averaged each score to give us our probability of relevance vector $\relevancies$ and covariance matrix $\Sigma$. The dependencies in this covariance matrix reflect the level of agreement between the different retrieval methods rather than direct correlations between the documents themselves i.e. similarly ranked documents will be positively correlated with one another. Finally, we selected those documents that had the top 30 mean relevance scores. These were then used by our algorithms to create rankings for two pages of search results with ten documents on each. 

We ranked these documents according to the baseline \emph{PRP} approach and also the already described IIR-PRP-MPS and DIR-MPS algorithms. We also investigated an interactive version of DIR-MPS (called IIR-MPS) that ranks the first page of results according to the \emph{PRP} and then optimizes a ranking for the second page of results by marginalizing over potential clicks using Eq.~(\ref{eq:expectedutility}). We also created a static version of DIR-MPS (called S-MPS) that removes feedback from $U_D$ entirely to give us the objective function $\utilityS(\actions_1, \relevancies) + \lambda \utilityS(\actions_2, \relevancies)$. We also investigated `perfect click' variants of the dynamic (DIR-MPS$^C$) and interactive (IIR-MPS$^C$) algorithms, where we interpret the hidden relevance labels as clicks on the first page of results, giving us the optimal observation setting and an upper bound on performance for the second page. 

To evaluate the quality of the rankings we measured MAP, NDCG and ERR for each page. For the DIR-MPS and IIR-MPS algorithms we actually generate an optimal 2nd page ranking for every click combination in our observation space, giving us different metrics scores for each. In these cases, the reported page two metric scores are averages over the page two scores for all click combination based rankings. This highlights an open area for research; the definition of evaluation metrics for \emph{DIR} that can take into account all of the potential rankings in a dynamic system. We also measure the session-based metrics sDCG (defined in Eq.~(\ref{eq:sDCG})) and sAP to evaluate performance over both pages, although it is worth noting that these metrics were designed for session search rather than multi page search. Finally, we also measure $\alpha-$NDCG~\cite{novelty-and-diversity}, Intent-Aware Precision (IA-Precision)~\cite{Agrawal:diversifying} and Intent-Aware ERR (ERR-IA)~\cite{ExpectedReciprocalRank} for scoring the diversity of rankings in the ClueWeb09 collection.

\setlength{\dbltextfloatsep}{0pt}

\begin{table*}[t]
  \centering
  \caption{NDCG, MAP and ERR scores for pages 1 and 2 of the search results and sAP and sDCG over both pages. Static, interactive and dynamic algorithms are grouped. The results shown are those for the optimal value of $\lambda$ in each collection, found by repeating the experiment for values in the range $[0,1]$. The maximum score for each metric on each page is given in boldface. A $^1$ or $^2$ indicates that the result is significantly better than the \emph{PRP} or \emph{IIR-PRP-MPS} baseline scores respectively using the Wilcoxon signed-rank test $(p < 0.05)$. }
  {\setlength{\extrarowheight}{1pt}%
    \begin{tabular}{|p{2cm}| l | c | c | c | c | c | c | c | c |}
    \cline{3-10}
\multicolumn{2}{c|}{}    & \multicolumn{3}{c|}{Page 1} & \multicolumn{3}{c|}{Page 2} & \multicolumn{2}{c|}{Both Pages} \\
\hline
Collection & Algorithm & NDCG & MAP   & ERR   & NDCG & MAP   & ERR   & sAP   & sDCG \\
\hline
\hline
{\multirow{7}{*}{\begin{minipage}{2cm}Web Track (WT10g) $\lambda=0.5$\end{minipage}}} & PRP   & \textbf{0.338} & \textbf{0.167} & \textbf{0.169} & 0.133 & 0.025 & 0.053 & 0.097 & 1.326 \\
  & IIR-PRP-MPS & 0.330 & 0.162 & 0.162 & 0.166 & 0.041 & 0.078 & 0.101 & 1.347 \\
    & S-MPS & 0.295 & 0.134 & 0.130 & 0.226 & \textbf{0.070} & 0.242$^{12}$ & 0.095 & 1.236 \\
    \cline{2-10}
    & IIR-MPS & \multirow{2}[4]{*}{\textbf{0.338}} & \multirow{2}[4]{*}{\textbf{0.167}} & \multirow{2}[4]{*}{\textbf{0.169}} & 0.125 & 0.025 & 0.103$^1$ & 0.097 & 1.291 \\
& IIR-MPS$^C$ &       &       &       & 0.154 & 0.040 & 0.151$^1$ & \textbf{0.102} & \textbf{1.353} \\
    \cline{2-10}
    & DIR-MPS & \multirow{2}[4]{*}{0.235} & \multirow{2}[4]{*}{0.091} & \multirow{2}[4]{*}{0.092} & 0.212$^1$ & 0.054$^1$ & 0.289$^{12}$ & 0.069 & 1.022 \\
    & DIR-MPS$^C$ &       &       &       & \textbf{0.230}$^1$ & 0.059 & \textbf{0.297}$^{12}$ & 0.072 & 1.027 \\
\hline
\hline
   \multirow{7}{*}{\begin{minipage}{2cm}Robust (AQUAINT) $\lambda = 0.5$\end{minipage}} & PRP   & 0.624 & \textbf{0.107} & \textbf{0.398} & 0.552 & 0.061 & 0.294 & \textbf{0.085} & 5.735 \\
& IIR-PRP-MPS & \textbf{0.629} & \textbf{0.107} & \textbf{0.398} & 0.514 & 0.052 & 0.288 & 0.083 & 5.680 \\
    & S-MPS & 0.608 & 0.096 & 0.388 & \textbf{0.595} & \textbf{0.066} & \textbf{0.887}$^{12}$ & 0.083 & \textbf{5.749} \\
    \cline{2-10}
   & IIR-MPS & \multirow{2}[4]{*}{0.624} & \multirow{2}[4]{*}{\textbf{0.107}} & \multirow{2}[4]{*}{\textbf{0.398}} & 0.519 & 0.050 & 0.737$^{12}$ & 0.081 & 5.543 \\
  & IIR-MPS$^C$ &       &       &       & 0.554 & 0.057 & 0.690$^{12}$ & 0.084 & 5.729 \\
    \cline{2-10}
    & DIR-MPS & \multirow{2}[3]{*}{0.548} & \multirow{2}[3]{*}{0.063} & \multirow{2}[3]{*}{0.304} & 0.575 & 0.065 & 0.656$^{12}$ & 0.065 & 4.921 \\
   & DIR-MPS$^C$ &       &       &       & 0.553 & 0.058 & 0.697$^{12}$ & 0.062 & 4.909 \\
    \hline
    \hline
    \multirow{7}{*}{\begin{minipage}{2cm}Diversity (ClueWeb09) $\lambda = 0.8$ \end{minipage}} & PRP   & 0.402$^2$ & \textbf{0.049}$^2$ & 0.199 & \textbf{0.476} & 0.052 & 0.265 & \textbf{0.051}$^2$ & \textbf{1.883}$^2$ \\
          & IIR-PRP-MPS & 0.384 & 0.046 & 0.193 & 0.468 & 0.051 & 0.257 & 0.048 & 1.808 \\
          & S-MPS & 0.388 & 0.041 & 0.193 & 0.465 & \textbf{0.054} & 0.358$^{12}$ & 0.049 & 1.787 \\
          \cline{2-10}
          & IIR-MPS & \multirow{2}[4]{*}{0.402$^2$} & \multirow{2}[4]{*}{\textbf{0.049}$^2$} & \multirow{2}[4]{*}{0.199} & 0.431 & 0.042 & 0.353$^{12}$ & 0.047 & 1.783 \\
          & IIR-MPS$^C$ &       &       &       & 0.436 & 0.042 & 0.345$^{12}$ & 0.047 & 1.787 \\
          \cline{2-10}
          & DIR-MPS & \multirow{2}[4]{*}{\textbf{0.451}$^{2}$} & \multirow{2}[4]{*}{0.047} & \multirow{2}[4]{*}{\textbf{0.238}$^{2}$} & 0.445 & 0.042 & \textbf{0.373}$^{12}$ & 0.046 & 1.859 \\
          & DIR-MPS$^C$ &       &       &       & 0.426 & 0.037 & 0.356$^{12}$ & 0.044 & 1.839 \\
    \hline
    \end{tabular}}
  \label{tbl:general-results}
\end{table*}

The results of our experiments are shown in Table~\ref{tbl:general-results}. For the Web Track and Robust collections, we observe that the 1st page losses of the dynamic techniques (when compared to the \emph{PRP} and \emph{IIR-PRP-MPS} baselines) are made up for by gains in the second page, significantly so on the WT10g dataset. Nonetheless, in these ad hoc ranking scenarios it is clear that the static \emph{PRP} and \emph{IIR-PRP} frameworks are still very effective.
 

\begin{table}[t]
  \centering
  \caption{$\alpha$-DCG, IA-Precision and ERR-IA scores for page 1 and 2 search results from the diversity track data. The maximum score for each metric on each page is given in boldface. A $^2$ indicates that the result is significantly better than the \emph{IIR-PRP-MPS} baseline score using the Wilcoxon signed-rank test $(p < 0.05)$. }
  \setlength\tabcolsep{3pt} 
    \begin{tabular}{|p{2.08cm}| p{0.9cm} | p{0.8cm} | p{0.8cm} | p{0.8cm} | p{0.8cm} | p{0.8cm} |}
    \cline{2-7}
\multicolumn{1}{c|}{}          & \multicolumn{3}{c|}{Page 1} & \multicolumn{3}{c|}{Page 2} \\
    \hline
    Algorithm & $\alpha$-DCG & IA-Prec & ERR-IA & $\alpha$-DCG & IA-Prec & ERR-IA\\
    \hline
    \hline
    PRP   & 0.360 & \textbf{0.083} & 0.239 & \textbf{0.420} & 0.085 & \textbf{0.294} \\
    IIR-PRP-MPS & 0.345 & 0.077 & 0.230 & 0.404 & 0.086 & 0.269 \\
    S-MPS & 0.352 & 0.078 & 0.233 & 0.417 & \textbf{0.089} & 0.280 \\
    \hline
    IIR-MPS & \multirow{2}[4]{*}{0.360} & \multirow{2}[4]{*}{\textbf{0.083}} & \multirow{2}[4]{*}{0.239} & 0.377 & 0.079 & 0.243 \\
    IIR-MPS$^C$ &       &       &       & 0.379 & 0.080 & 0.236 \\
    \hline
    DIR-MPS & \multirow{2}[4]{*}{\textbf{0.403$^2$}} & \multirow{2}[4]{*}{0.082} & \multirow{2}[4]{*}{\textbf{0.270}} & 0.400 & 0.079 & 0.264 \\
    DIR-MPS$^C$  &       &       &       & 0.386 & 0.077 & 0.254 \\
    \hline
    \end{tabular}%
  \label{tbl:diversity-results}%
\end{table}%

We see different results with the diversity track data. The metric scores for this data in Table~\ref{tbl:general-results} were calculated using relevance judgments from all subtopics. We see the opposite relationship between page scores here, with DIR-MPS having higher scores for the 1st page and losses in the second (except for ERR which is significantly improved across both pages). We see further evidence of this with the diversity metric scores in Table~\ref{tbl:diversity-results}, where it is clear that diversification is occurring in the first page and less so in the second. This backs up our intuition (in Fig.~\ref{fig:dir-example-d}) that a dynamic technique will initially diversify results to improve future rankings, and also helps explain the losses in performance of the 1st page in the other datasets (which do not have subtopic relevance judgments). The reduced diversity of the 2nd page indicates that it is more tightly focused on the user's subtopic preference. 

We observe that the diversity task is more suited as an application of the dynamic IR framework. This is evidenced by the optimal settings for $\lambda$ in each collection. For the ad hoc ranking task in the WT10g and AQUAINT collections, the setting for $\lambda$ gives greater weight to the 1st page of results, rewarding immediately effective rankings. Whereas in the diversity task, the utility of the 2nd page has a larger effect on the overall utility, encouraging diversity. 

Further to this, the interactive variant scored highly with \emph{session based metrics} on the Robust dataset, but otherwise the static techniques were optimal, even for the diversity task. This may partly be due to the application of a session-based metric to the multi page scenario and also the inability of the metric to take into account the user interaction. Finally, we also see that the `perfect click' variants generally outperform their counterparts (except over the diversity data), indicating that the 2nd page ranking can be improved when high quality clicks are observed. 

In summary, by its nature the \emph{DIR} approach to multi-page search places greater emphasis on different stages of the search task. We find that this may not be suited to all search environments i.e. ad hoc search. In such cases the static approaches can be more effective. Nonetheless, the dynamic IR framework has other desirable properties such as the diversification and personalization of results over time.

\section{Related Work}


Throughout this paper we have presented the dynamic IR theory within the context of the surrounding literature, so in this section we cover those areas of the related work not already discussed.


For instance, the settings of the components in the \emph{DIR} framework for multi page search cover a wide area of research in IR. Firstly, the examination hypothesis model used is just one of a number of probabilistic click models that could have been employed, including the click-chain model~\cite{Guo:2009:CCM:1526709.1526712} and even a POMDP-based model~\cite{Wang:2010:ISB:1718487.1718514}. Other path-based discount functions have been explored in the literature \cite{White:2005:EIF:1080343.1080347} as well as other multi-stage utilities and metrics such as Time-Based Gain~\cite{Smucker:2012:TCE:2348283.2348300}. Related work on using Markov chains to measure the utility of rankings at each time step is a potential method for evaluating \emph{DIR} problems~\cite{Ferrante:2014:IUM:2600428.2609637}. Further to this, the identification of the dynamic IR framework as a POMDP raises the possibility of using established techniques such as the Witness algorithm~\cite{Littman:1994:WAS:864404} to find optimal action policies. Also, the performance of the \emph{PRP} under results diversification is well-reported in the static frameworks quantum-PRP~\cite{quantum-prp} and the portfolio theory of IR~\cite{Wang09portfoliotheory}. 

The concept of evaluating for retrieval utility rather than relevance was proposed by Cooper in 1973~\cite{ASI:ASI4630240204} and is extended to all the frameworks discussed in this paper, where we aim to maximize some utility function that balances the costs and benefits of an IR system's actions. Other work in this area includes Azzopardi's~\cite{Azzopardi:2014:MIE:2600428.2609574} work on economic models, which is itself an extension of the \emph{IIR-PRP}, and also the work of Mostafa et al. \cite{simulation-studies} who had a similar motivation to this work, where they defined a framework for running user simulations for information filtering, itself a \emph{DIR} problem. 

Other than the \emph{PRP} and \emph{IIR-PRP}, the closest related works to this one are the following: The application of a POMDP to multi page search~\cite{Jin:2013:IES:2488388.2488446}, from which many aspects of our experiments in this paper are derived, including the problem setting and the probability of relevance distribution. This paper extends their formulation to a general one applicable to other \emph{DIR} problems and explores a different setting for the static utility, observation likelihood and discount, while also linking to static and interactive techniques. Our experimental time horizon setting of $T=2$ is based on the optimal results found in their work.
Finally, the work on defining the elements of a POMDP in session search~\cite{grace-ecir} is a close relation to this work, though focusing more on the testing of particular settings of \emph{DIR} components in the session search scenario rather than explicitly gaining an understanding of the framework and components themselves. Nonetheless, their work contains an evaluation of algorithms that fall under the \emph{DIR} framework including one similar to DIR-MPS.

This work differs from the literature in that: 1) ours is the first work to define the characteristics that distinguish dynamic IR from the other theoretical IR frameworks, 2) our utility is the generalization of many existing ranking utilities and incorporates many elements of IR research such as click models, and 3) we confirm the effectiveness of ours and the static frameworks in different scenarios in our experiments.

\section{Conclusion}

In this paper we have established a theoretical framework for \emph{Dynamic Information Retrieval}. By contrasting with \emph{static} and \emph{interactive} frameworks, we found three characteristics that define dynamic IR systems; user feedback, temporal dependency and an overall goal. This motivated the derivation of our dynamic utility function $\utilityD$, which has its roots in the POMDP formulation. The components of this utility can be directly implemented using elements from existing research which we apply in the DIR-MPS algorithm, an example instantiation designed for the multi page search problem. Our experiments confirm that in this scenario, one of the effects of optimizing for $\utilityD$ is the diversification of search results. Otherwise, we also demonstrate that for other scenarios the \emph{PRP} and \emph{IIR-PRP} frameworks are still effective. 

Like the \emph{PRP} and \emph{IIR-PRP}, our framework defines certain functional parameters but does not definitively  specify how to set them. Instead, this work is a point of reference that can be used for the development of specialized models and algorithms applied to \emph{DIR} problems. Through our application we were able to consider the limitations of the \emph{DIR} framework, a result of which is the approximations used in the DIR-MPS algorithm. 
Finally, the derivation of a stationary solution to the \emph{DIR} Markovian model is an intended future goal and would be an important result to come from this work.

\balance
\bibliographystyle{acm}
\begin{scriptsize}
\bibliography{refs}

\begin{thebibliography}{10}
\vfill
\bibitem{Agrawal:diversifying}
{\sc Agrawal, R., Gollapudi, S., Halverson, A., and Ieong, S.}
\newblock Diversifying search results.
\newblock WSDM '09, ACM, pp.~5--14.

\bibitem{Azzopardi:2014:MIE:2600428.2609574}
{\sc Azzopardi, L.}
\newblock Modelling interaction with economic models of search.
\newblock SIGIR '14, ACM, pp.~3--12.

\bibitem{Bellman:2003:DP:862270}
{\sc Bellman, R.~E.}
\newblock {\em Dynamic Programming}.
\newblock Dover Publications, 2003.

\bibitem{ExpectedReciprocalRank}
{\sc Chapelle, O., Metlzer, D., Zhang, Y., and Grinspan, P.}
\newblock Expected reciprocal rank for graded relevance.
\newblock CIKM '09, ACM, pp.~621--630.

\bibitem{Chen:2006:LMP:1148170.1148245}
{\sc Chen, H., and Karger, D.~R.}
\newblock Less is more: Probabilistic models for retrieving fewer relevant
  documents.
\newblock SIGIR '06, ACM, pp.~429--436.

\bibitem{novelty-and-diversity}
{\sc Clarke, C.~L., Kolla, M., Cormack, G.~V., Vechtomova, O., Ashkan, A.,
  B\"{u}ttcher, S., and MacKinnon, I.}
\newblock Novelty and diversity in information retrieval evaluation.
\newblock SIGIR '08, ACM, pp.~659--666.

\bibitem{Cleverdon68}
{\sc Cleverdon, C., and Kean, M.}
\newblock Factors determining the performance of indexing systems.
\newblock Aslib Cranfield Research Project, Cranfield, England, 1968.

\bibitem{cooper1971inadequacy}
{\sc Cooper, W.~S.}
\newblock The inadequacy of probability of usefulness as a ranking criterion
  for retrieval system output.
\newblock {\em University of California, Berkeley\/} (1971).

\bibitem{ASI:ASI4630240204}
{\sc Cooper, W.~S.}
\newblock On selecting a measure of retrieval effectiveness.
\newblock {\em Journal of the American Society for Information Science 24}, 2
  (1973), 87--100.

\bibitem{clickmodels}
{\sc Craswell, N., Zoeter, O., Taylor, M., and Ramsey, B.}
\newblock An experimental comparison of click position-bias models.
\newblock WSDM '08, ACM, pp.~87--94.

\bibitem{Ferrante:2014:IUM:2600428.2609637}
{\sc Ferrante, M., Ferro, N., and Maistro, M.}
\newblock Injecting user models and time into precision via {Markov} chains.
\newblock SIGIR '14, ACM, pp.~597--606.

\bibitem{DBLP:journals/ir/Fuhr08}
{\sc Fuhr, N.}
\newblock A probability ranking principle for interactive information
  retrieval.
\newblock {\em Inf. Retr. 11}, 3 (2008), 251--265.

\bibitem{Guo:2009:CCM:1526709.1526712}
{\sc Guo, F., Liu, C., Kannan, A., Minka, T., Taylor, M., Wang, Y.-M., and
  Faloutsos, C.}
\newblock Click chain model in web search.
\newblock WWW '09, ACM, pp.~11--20.

\bibitem{Hersh:2000:BUE:345508.345539}
{\sc Hersh, W., Turpin, A., Price, S., Chan, B., Kramer, D., Sacherek, L., and
  Olson, D.}
\newblock Do batch and user evaluations give the same results?
\newblock SIGIR '00, ACM, pp.~17--24.

\bibitem{Jambor:2012:UCT:2187836.2187839}
{\sc Jambor, T., Wang, J., and Lathia, N.}
\newblock Using control theory for stable and efficient recommender systems.
\newblock WWW '12, ACM, pp.~11--20.

\bibitem{Jansen:2006:WSW:1138797.1138813}
{\sc Jansen, B.~J., and Spink, A.}
\newblock How are we searching the world wide web?: A comparison of nine search
  engine transaction logs.
\newblock {\em Inf. Process. Manage. 42}, 1 (2006), 248--263.

\bibitem{Jansen:2005:TCA:1059467.1059470}
{\sc Jansen, B.~J., Spink, A., and Pedersen, J.}
\newblock A temporal comparison of {AltaVista} web searching: Research
  articles.
\newblock {\em J. Am. Soc. Inf. Sci. Technol. 56}, 6 (2005), 559--570.

\bibitem{Jarvelin:2008:DCG:1793274.1793280}
{\sc J\"{a}rvelin, K., Price, S.~L., Delcambre, L. M.~L., and Nielsen, M.~L.}
\newblock Discounted cumulated gain based evaluation of multiple-query {IR}
  sessions.
\newblock ECIR'08, Springer-Verlag, pp.~4--15.

\bibitem{Jin:2013:IES:2488388.2488446}
{\sc Jin, X., Sloan, M., and Wang, J.}
\newblock Interactive exploratory search for multi page search results.
\newblock In {\em WWW '13\/} (2013), pp.~655--666.

\bibitem{Joachims:2007:EAI:1229179.1229181}
{\sc Joachims, T., Granka, L., Pan, B., Hembrooke, H., Radlinski, F., and Gay,
  G.}
\newblock Evaluating the accuracy of implicit feedback from clicks and query
  reformulations in web search.
\newblock {\em ACM Trans. Inf. Syst. 25}, 2 (2007).

\bibitem{Kanoulas:2011:EMS:2009916.2010056}
{\sc Kanoulas, E., Carterette, B., Clough, P.~D., and Sanderson, M.}
\newblock Evaluating multi-query sessions.
\newblock SIGIR '11, ACM, pp.~1053--1062.

\bibitem{Kim:2013:UPI:2541176.2505663}
{\sc Kim, J.~Y., Cramer, M., Teevan, J., and Lagun, D.}
\newblock Understanding how people interact with web search results that change
  in real-time using implicit feedback.
\newblock CIKM '13, ACM, pp.~2321--2326.

\bibitem{Littman:1994:WAS:864404}
{\sc Littman, M.~L.}
\newblock The witness algorithm: Solving partially observable markov decision
  processes.
\newblock Tech. rep., 1994.

\bibitem{grace-ecir}
{\sc Luo, J., Zhang, S., Dong, X., and Yang, H.}
\newblock Designing states, actions, and rewards for using {POMDP} in session
  search.
\newblock In {\em Advances in Information Retrieval}, vol.~9022 of {\em Lecture
  Notes in Computer Science}. Springer International Publishing, 2015,
  pp.~526--537.

\bibitem{simulation-studies}
{\sc Mostafa, J., Mukhopadhyay, S., and Palakal, M.}
\newblock Simulation studies of different dimensions of users' interests and
  their impact on user modeling and information filtering.
\newblock {\em Information Retrieval 6}, 2 (2003), 199--223.

\bibitem{JD:1977:Robertson:PRP}
{\sc Robertson, S.~E.}
\newblock {The Probability Ranking Principle in IR}.
\newblock {\em Journal of Documentation 33}, 4 (1977), 294--304.

\bibitem{Rocchio}
{\sc Rocchio, J.}
\newblock {\em {Relevance Feedback in Information Retrieval}}.
\newblock 1971, pp.~313--323.

\bibitem{DBLP:journals/arist/Ruthven08}
{\sc Ruthven, I.}
\newblock Interactive information retrieval.
\newblock {\em ARIST 42}, 1 (2008), 43--91.

\bibitem{my-ijr-work}
{\sc Sloan, M., Yang, H., and Wang, J.}
\newblock A term-based methodology for query reformulation understanding.
\newblock {\em Information Retrieval Journal 18}, 2 (2015), 145--165.

\bibitem{Smucker:2012:TCE:2348283.2348300}
{\sc Smucker, M.~D., and Clarke, C.~L.}
\newblock Time-based calibration of effectiveness measures.
\newblock SIGIR '12, ACM, pp.~95--104.

\bibitem{Sondik:1978}
{\sc Sondik, E.}
\newblock The optimal control of partially observable markov processes over the
  infinite horizon: Discounted cost.
\newblock {\em Operations Research 26}, 2 (1978), 282--304.

\bibitem{Wang09portfoliotheory}
{\sc Wang, J., and Zhu, J.}
\newblock Portfolio theory of information retrieval.
\newblock SIGIR' 09, ACM, pp.~115--122.

\bibitem{WANG:2010:SAO:1835449.1835489}
{\sc Wang, J., and Zhu, J.}
\newblock On statistical analysis and optimization of information retrieval
  effectiveness metrics.
\newblock SIGIR '10, pp.~226--233.

\bibitem{Wang:2010:ISB:1718487.1718514}
{\sc Wang, K., Gloy, N., and Li, X.}
\newblock Inferring search behaviors using partially observable markov {(POM)}
  model.
\newblock WSDM '10, ACM, pp.~211--220.

\bibitem{white2009exploratory}
{\sc White, R., White, R., and Roth, R.}
\newblock {\em Exploratory Search: Beyond the Query-Response Paradigm}.
\newblock Synthesis Lectures on Information Concepts, Retrieval, and Services
  Series. Morgan \& Claypool, 2009.

\bibitem{White:2005:EIF:1080343.1080347}
{\sc White, R.~W., Ruthven, I., Jose, J.~M., and Rijsbergen, C. J.~V.}
\newblock Evaluating implicit feedback models using searcher simulations.
\newblock {\em ACM Trans. Inf. Syst. 23}, 3 (2005), 325--361.

\bibitem{yangLad}
{\sc Yang, Y., and Lad, A.}
\newblock Modeling expected utility of multi-session information distillation.
\newblock In {\em Advances in Information Retrieval Theory}, vol.~5766 of {\em
  Lecture Notes in Computer Science}. Springer Berlin Heidelberg, 2009,
  pp.~164--175.

\bibitem{quantum-prp}
{\sc Zuccon, G., Azzopardi, L., and van Rijsbergen, K.}
\newblock The quantum probability ranking principle for information retrieval.
\newblock In {\em Advances in Information Retrieval Theory}, vol.~5766.
  Springer Berlin Heidelberg, 2009, pp.~232--240.

\end{thebibliography}
\end{scriptsize}

\balancecolumns
\end{document}